\documentclass[aps,twocolumn,showpacs]{revtex4}
\usepackage{graphicx}

\begin{document}
\title{How to accurately measure the mobility and viscosity of two-dimensional carriers?}
\author{M.V. Cheremisin}
\affiliation{A.F.Ioffe Physical-Technical Institute, 194021, St.Petersburg, Russia}
\author{M.A. Zudov}
\affiliation{School of Physics and Astronomy, University of Minnesota, Minneapolis, Minnesota 55455, USA}
\date{\today}
\begin{abstract}
Two different methods of \emph{metrological} accuracy are proposed to determine the mobility and viscosity of ultra clean two-dimensional electron liquids. The  experimental data analysis is based on the Gurzhi hydrodynamic model under no-slip boundary conditions without preliminary assumptions about carrier scattering times. The applicability of no-slip boundary conditions has been proven. A Hall bar with several conducting channels of different widths in a zero magnetic field and a sample with a single channel in a perpendicular field are considered. In both cases, it was possible to accurately isolate the ohmic part of the total measured resistance and, then find the exact mobility and viscosity of the charge liquid. The extracted e-e scattering time is extremely close to that obtained by other experimental group for transport measurements of superballistic point contact. At low temperatures the e-e scattering time demonstrates a stronger dependence compare to $1/T^{2}$ behavior predicted by Fermi liquid theory. We propose both methods as powerful tools for viscometry and finding the mobility of two-dimensional systems.
\end{abstract}
\maketitle

\section{Introduction}
\label{Introduction}
In the 1960s, R.N. Gurzhi substantiated\cite{Gurzhi63,Gurzhi68} the applicability of the hydrodynamic approach to the description of strongly interacting electrons in metals. Unfortunately, such behavior has not been observed due to insufficient purity of ordinary metals. Recently, the emergence of high mobility two-dimensional (2D) systems, graphene and van der Waals heterostructures, has generated interest in the behavior of charged liquids. In general, the description of viscous charged liquids is based on the solution of the Boltzmann kinetic equation\cite{Muller09,Andreev11,Briskot15, Alekseev16,Levitov16,Scaffidi17}. However, in most problems it is sufficient to use the Navier-Stokes hydrodynamic equations, which can be derived\cite{Muller09} from the kinetic equation. Following the general Navier-Stokes formalism, two regimes of hydrodynamic flow are distinguished. The linear laminar mode has been observed in the form of Knudsen\cite{Molenkamp94,Molenkamp95} and Poiseuille\cite{Mani13,Sulpizio19,Ku20,Wang22} flows. Then, successful vsisualization of nonlinear turbulent flow regime of charged fluid revealed the presence of vortices\cite{Aharon-Steinberg22} and counterflows\cite{Bandurin16,Levin18,Bandurin18,Gupta21}. The article is devoted to a detailed analysis of the ordinary Poiseuille flow in a channel of arbitrary width in the absence\cite{Mani13,Wang22} and presence\cite{Dai10,Hatke11,Bockhorn11,Hatke12,Bockhorn14,Zudov14,Gusev18,Gusev18+,Horn21,Levin25} of a perpendicular magnetic field. We have identified a number of inaccuracies and simplifications in previous studies\cite{Torre15,Alekseev16,Kiselev19} that are critical to our goal of accurate mobility and viscosity finding.

\section{Gurzhi flow model}
\label{Gurzhi flow model}
For clarity, we will further consider a rectangular sample in the x-y plane, the length of which in the x direction is significantly greater than the width, $w$. The smallest dimension, namely the width, defines an important length scale for the problem under study. It is generally accepted that electron fluidity occurs when the electron-electron(e-e) scattering length, $l_{ee}$, is smaller than the typical sample size, $w$, and the length, $l_{p}$, describing the momentum transfer from electrons to the lattice. It is stated that the e-e scattering length uniquely determines the viscosity of the electron fluid. Recently, the importance of the static disorder contribution to viscosity has been rethoughted\cite{Alekseev16,Zakharov21}. It is argued that the viscosity of a fluid depends on relaxation time $\tau_{2}$ of the second angular harmonic of the distribution function. The respective length $l_{2}=\tau_{2}v_{F}$ reads
\begin{equation}
l_{2}^{-1}=l_{ee}^{-1}+l_{d}^{-1},
\label{Length_2}\\
\end{equation}
where $v_{F}$ is the Fermi velocity, $l_{d}$ is the length associated with static disorder. At low temperatures, the rate of e-e collisions slows down\cite{Pomeranchuk50} such that $1/l_{ee}\sim AT^{2}$, which makes the static disorder term in Eq.(\ref{Length_2}) particularly important. In connection with the above, the correct fluidity condition yields
\begin{equation}
l_{2}\ll w^{>}_{<} l_{p}.
\label{Fluidity_Condition}\\
\end{equation}

In perpendicular magnetic field $B=B_{z}$, the linearized Navier-Stokes hydrodynamic equation has the form:
\begin{equation}
\frac{\partial {\bf V}}{\partial t}=\eta_{xx}\triangle{\bf V}+[(\eta_{yx}\triangle{\bf V}+\omega_{c}{\bf V})\times {\bf e}_{z}]+\frac{e{\bf E}}{m}-\frac{{\bf V}}{\tau_{p}}.
\label{Navier-Stokes}\\
\end{equation}
Here, the charged liquid is assumed to be incompressible and the condition $\textrm{div}{\bf V}=0$ is satisfied. The $\bf{V}$ and $\bf{E}$ are the in-plane flow velocity and the electric field, respectively, $\omega_{c}$ is the cyclotron frequency. Then, $\eta_{xx}=\frac{\eta}{\sqrt{1+(2 l_{2}/r_{c})^{2}}}$ and $\eta_{yx}=\frac{2 l_{2}}{r_{c}}\eta_{xx}$ are the longitudinal and transverse components\cite{Chapman53} of the viscosity tensor, respectively. Finally, $\eta=v_{F}l_{2}/4$ is the 2D viscosity in zero magnetic field\cite{Alekseev16}, $r_{c}=v_{F}/\omega_{c}$ is the cyclotron radius.

Let the electric field $E=E_{x}$ be applied along $x$-axis. For a steady flow, the Eq.(\ref{Navier-Stokes}) can be written in a form
\begin{equation}
\eta_{xx}\frac{d^{2}V}{dy^{2}}+\frac{e}{m}E_{x}-\frac{V}{\tau_{p}}=0,
\label{Gurzhi_Flow}\\
\end{equation}
that allows one to determine the transverse profile of the longitudinal velocity $V=V_{x}(y)$.

\subsection{Boundary conditions}
\label{Boundary conditions}
The problem of the boundary conditions is of special interest. As discussed by Maxwell\cite{Maxwell1867}, the
boundary condition can, in general, be written as
\begin{equation}
V|_{\pm w}=\zeta\textbf{n} \nabla V|_{\pm w},
\label{Maxwell}\\
\end{equation}
where $\zeta$ is the so-called slip length, \textbf{n} is the normal to the boundary. The Eq.(\ref{Maxwell}) describes no-slip $V=0$ and no-stress $\nabla V=0$ limiting cases for the slip length $\zeta=0$ and $\zeta=\infty$, respectively. With the help of  Eqs.(\ref{Gurzhi_Flow}),(\ref{Maxwell}) one may readily find the transverse profile of the flow velocity and, then the average flow velocity:
\begin{equation}
\bar{V}=V_{D}\left [ 1-\frac{1}{\xi_{B}(\coth{\xi_{B}}+2\zeta \xi_{B}/w)} \right ],
\label{Gurzhi_Maxwell_Resistivity}\\
\end{equation}
where $V_{D}=\frac{e\tau_{p}}{m} E_{x}$ is the drift velocity. Here, we have introduced the dimensionless Gurzhi parameter $\xi_{B}=\xi\sqrt{1+(2 l_{2}/r_{c})^{2}}$, where $\xi=w/l_{G}$ denotes its value in zero magnetic field. Further, $l_{G}=\sqrt{l_{2}l_{p}}$ is the so-called Gurzhi length which is another important length scale of the problem under study, in addition to the sample width. In fact, the smaller of the lengths $l_{G}$ or $w$ determines the scale of the spatial change of the flow velocity in the transverse direction. 

We emphasize that the criterion $l_{2}<w$ given by Eq.(\ref{Fluidity_Condition}) for applicability of the hydrodynamic approach provides a lower limit for the Gurzhi parameter. Indeed, for high-mobility 2D liqud, i.e. when $l_{2}\ll l_{p}$, one obtains the condition $\xi>\xi_{min}$, where $\xi_{min}=\sqrt{l_{2}/l_{p}} \ll 1$ is the the minimum value of the Gurzhi parameter for which the hydrodynamic approach is valid.

Surprisingly, a previously known\cite{Torre15} Eq.(\ref{Gurzhi_Maxwell_Resistivity}) sheds light on the issue of boundary conditions. To confirm this, let us estimate the term containing the slip length in the denominator of the Eq.(\ref{Gurzhi_Maxwell_Resistivity}). For a realistic case of diffusion scattering of carriers at boundaries, the slip length is close\cite{Pellegrino2017,Kiselev19} to length $l_{2}$ which determines the viscosity of the liquid. At zero magnetic field, i.e. when $\xi_{B}=\xi$, the slip term $2\zeta \xi/w\simeq 2\xi_{min}\ll 1$ in the denominator can be safely neglected compared to greater term $\coth{\xi}>1$. Thus, we conclude that the no-slip boundary condition is admissible at zero magnetic field. Of particular interest is the case of a non-zero magnetic field. At strong fields, i.e. when $2l_{2}/r_{c}\gg 1$, we have $\coth{\xi_{B}} \geq 1$, whereas the slip term becomes comparable to unity $2\zeta \xi_{B}/w \sim 1$ at certain cyclotron radius $r_{c}\simeq 4 \xi_{min} l_{2}$. This condition determines the critical magnetic field above which the consistent accounting of the finite slip length is necessary. For typical scattering lengths $l_{2}\sim 2\mu m$ and $l_{p}\sim 200 \mu m$ associated with high-mobility carriers in GaAs/AlGaAs system we estimate the cyclotron radius $r_{c}=0.8 \mu m$ and, in turn, the critical magnetic field $B\simeq 0.11$T. For channel width $w=50 \mu m$ we obtain $\xi_{B}\simeq 12.7$, which gives a contribution of a few percent to the Drude result $\bar{V}=V_{D}$ from the second term in the Eq.(\ref{Gurzhi_Maxwell_Resistivity}). The presented analysis confirms the validity of using no-slip boundary conditions in the general case.

By setting $\zeta=0$ in the Eq.(\ref{Gurzhi_Maxwell_Resistivity}), we reproduce the result of Ref.\cite{Alekseev16} for longitudinal magnetoresistivity:
\begin{equation}
\rho_{B}=\rho_{D}F(\xi_{B}),
\label{Gurzhi_Resistivity}\\
\end{equation}
where $F(\xi)=\frac{1}{1-\tanh{\xi}/\xi}$ is an auxiliary function, $\rho_{D}=\frac{m v_{F}}{ne^{2} l_{p}}$ is the Drude resistivity, $n$ is the 2D carrier density.

In practice, the exact values of the scattering lengths $l_{p}, l_{2}$ are unknown both theoretically and experimentally. Therefore, determining the scattering lengths is an ambitious task. Next we will consider two methods for determining both lengths using data from two different experimental setups.

\section{Transport in multiple-in-a-row Hall bars at zero magnetic field.}
\label{Transport in multiple-in-a-row Hall bars at zero magnetic field.}

\begin{figure}[tbp]
\begin{center}\leavevmode
\includegraphics[width=0.8\linewidth]{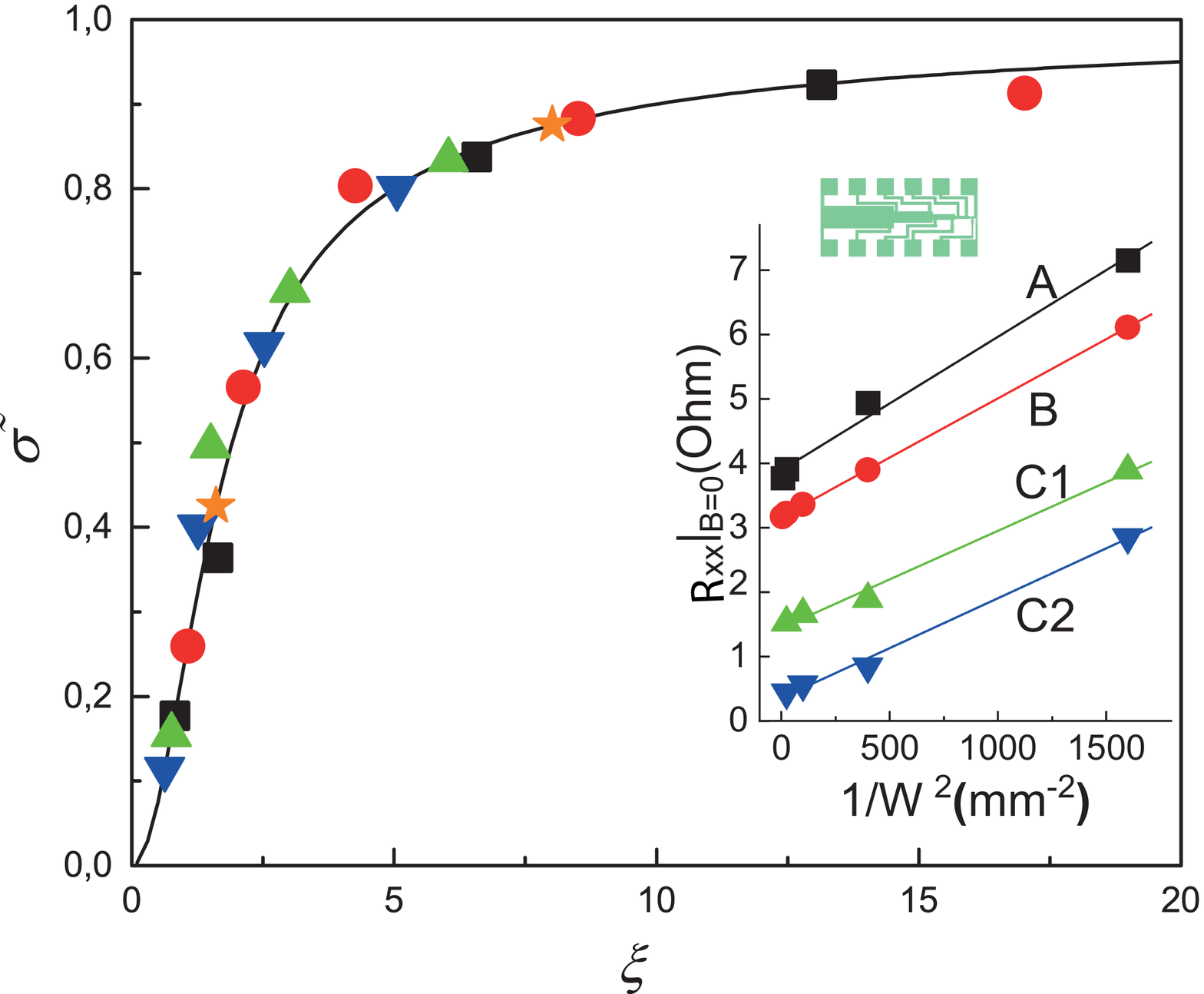} \caption[]{\label{Fig1} Inset: Original data reproduced from Ref.\cite{Wang22} for samples C2 and C1,A,B (shifted up by 1$\Omega$ successively). The figure shows the in-plane geometry of a five-row Hall rod sample. Main panel: Redrawing the data in the inset as dimensionless conductivity $\tilde{\sigma}=\rho_{D}/\rho$ vs Gurzhi parameter $\xi$. Asterisks indicate data from Ref.\cite{Horn21} for two separate samples of different widths 20$\mu m$ and 100 $\mu m$.}
\end{center}
\end{figure}

Let us first consider charge transfer in a zero magnetic field by substituting $\xi_{B}\rightarrow \xi$ and $\rho_{B}\rightarrow \rho$ into the Eq.(\ref{Gurzhi_Resistivity}). For a massive sample $\xi \gg 1$ hydrodynamic effects are insignificant. The resistivity has the usual form $\rho=\rho_{D}$, known in the Drude model. In the opposite case of a narrow channel $\xi \leq 1$, the expansion of the right-hand side of the equation (\ref{Gurzhi_Resistivity}) takes the form
\begin{equation}
\rho=\rho_{D}(6/5+3/\xi^{2}).
\label{Gurzhi_Resistivity_expansion}\\
\end{equation}
Moreover, with further narrowing of the channel $\xi\rightarrow \xi_{min} \ll 1$ the second term in the Eq.(\ref{Gurzhi_Resistivity_expansion}) prevails, determining the so-called viscous resistivity $\rho_{\eta}=\frac{m}{ne^{2}} \frac{3v_{F} l_{2}}{w^{2}}$, far superior to ohmic one. It is noteworthy that the relation $\rho_{\eta}\sim l_{2}/w^{2}$, valid for a narrow channel, allows us to find both scattering lengths $l_{p}, l_{2}$.
Indeed, let us imagine a 2D system (see Fig.\ref{Fig1}, inset), consisting of several Hall bars arranged in a row\cite{Mani13} with channels of different widths. The resistivity of each channel can be plotted as a function of the inverse square of the width of that channel $\rho(w^{-2})$. The part of the above dependence related to narrow channel can be extrapolated to the region of wide channels when $w \rightarrow \infty$. Extrapolation yields a resistivity cutoff 1.2 times the Drude value. Additionally, the slope of the narrow-channel part of the dependence $\rho(w^{-2})$ uniquely determines the viscosity and, in turn, the scattering length $l_{2}$. Note that this method has been extensively discussed in the literature\cite{Wang22}. Curiously, a currently known way is based on the empirical expression $\rho=\rho_{D}(1+3/\xi^{2})$, first proposed in Ref.\cite{Alekseev16} and intensively used by various experimental groups\cite{Levin18,Gusev18,Gusev18+,Levin25}. Although the empirical expression is close to exact Eq.(\ref{Gurzhi_Resistivity}), it is completely useless regarding extracting scattering lengths. Namely, the narrow-channel extrapolation $\xi\rightarrow \infty$ leads, according to the empirical formula, to cutoff value which would be incorrectly taken as the Drude resistance. Recall again that Eq.(\ref{Gurzhi_Resistivity_expansion}) gives the correct Drude resistance value which is 1.2 times less than the approximation cutoff value.

Generally speaking, the efficiency of the extrapolation method depends on the number $\geq 3$ of channels being studied, at least two of which must be narrow. However, it is quite difficult to verify a priori the fulfillment of the narrow channel condition $\xi \ll 1$ for an arbitrary sample. In what follows, we will avoid using extrapolation methods with the goal of accurately determining the transport properties of a 2D system.

We emphasize that use of multiple-in-row Hall bars is not necessary. Next, we propose a simple scheme using at least two different channels with a known width ratio $\kappa$. It is assumed that mobility, density and temperature should be the same for both channels. The resistivity of the narrow (first) and wide (second) channels is given by the following expressions:
\begin{eqnarray}
\rho_{1}=\rho_{D}F(\xi),
\label{First_channels}\\
\rho_{2}=\rho_{D}F(\kappa \xi),
\nonumber
\end{eqnarray}
where $w_{2}/w_{1}=\kappa >1$ is the ratio of channel widths. Solving the system of Eq.(\ref{First_channels}) allows one to find the Gurzhi parameter for the both channels, and then the Drude resistivity $\rho_{D}$ and, finally, the scattering lengths $l_{p}, l_{2}$. For samples with multiple channels, the scattering lengths are adjusted in order to best agree the experimental data with Eq.(\ref{Gurzhi_Resistivity}).

\begin{table}[h]
\begin{center}
\label{Data_collect} \caption{2D liquid parameters under Ref.\cite{Wang22}(Fig.\ref{Fig1},inset) and Ref.\cite{Horn21}}
\begin{tabular} {@{}*{10}{l}}
 \hline\hline &$T$& $n\cdot 10^{11}$ & $w$ & $\rho_{D}$ & $\mu\cdot 10^{6}$ & $l_{p}$ & $l_{2}$ & $\nu$ & Ref.\\
 sample &[K]&[cm$^{-2}$] &[$\mu m$] & [$\Omega$] & [$\frac{cm^{2}}{Vs}$] & [$\mu m$] & [$\mu m$] & [$\frac{m^{2}}{s}$] & \\
 \hline  $\fbox{}$ - A &0.3& 2.01 & 25-400 & 0.73 & 39.6 & 285 & 3.3 & 0.16 & \cite{Wang22}\\
         $\bigcirc$ - B &0.3& 2.62 & 25-400 & 1.08 & 22.0 & 187  & 2.9  & 0.16 & \cite{Wang22}\\
         $\bigtriangleup$ - C1 &0.3& 3.64 & 25-400 & 0.44 & 38.7 & 340 & 3.2 & 0.21 & \cite{Wang22}\\
         $\bigtriangledown$ - C2 &0.3& 4.20 & 25-400 & 0.32 & 46.1 & 402 & 3.9 & 0.28 & \cite{Wang22}\\
         $\star$ &0.8& 2.50 & 20;100 & 2.89 & 8.70 & 71 & 2.1 & 0.12 & \cite{Horn21}\\
\hline\\
\end{tabular}
\end{center}
\end{table}

As an example, we reproduced in Fig.\ref{Fig1}, inset the experimental data of Ref.\cite{Wang22}. Four different five-in-a-row Hall bar samples had the channel widths 25,50,100,200 and 400$\mu m$ respectively. It is clearly visible that the experimental data that fell into the wide channel region are very concentrated. To avoid this problem, the same data can be rearranged by plotting the conductivity versus channel width. Then, for each sample we fit the conductivity data. For each sample we find the Gurzhi parameter for the narrowest and, in turn, the all remaining channels. Then, for each sample we find correct Drude resistivity $\rho_{D}$ and, then compute and put in Table the scattering lengths $l_{p},l_{2}$. Fig.\ref{Fig1} shows the original experimental data reconstructed as a dependence of the reduced conductivity $\tilde{\sigma}=\rho_{D}/\rho$ on the Gurzhi parameter. The reference dependence $\tilde{\sigma}=1/F(\xi)$ given by Eq.(\ref{Gurzhi_Resistivity}) is plotted as a guide for eyes. To demonstrate the capabilities of the method, we analyzed the data\cite{Horn21} obtained for two separate samples with different channel widths of 20 and 100$\mu$, respectively. Except width, these samples had otherwise identical parameters. The result is shown in Fig.\ref{Fig1}, and the calculated scattering lengths are given in the Table.

\begin{figure}[tbp]
\begin{center}\leavevmode
\includegraphics[width=0.8\linewidth]{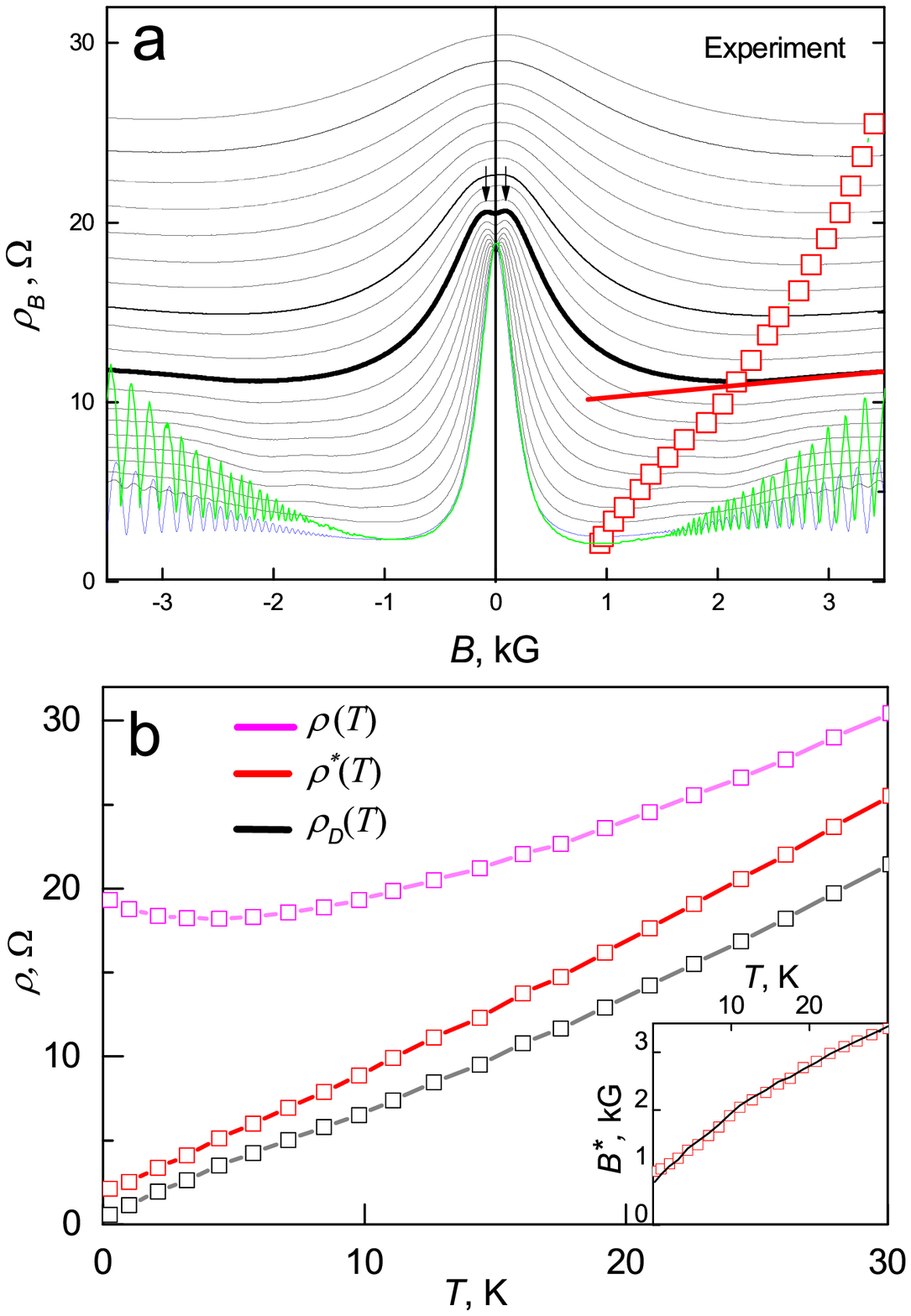} \caption[]{\label{Fig2} a) Magnetoresistivity data under Ref.\cite{Zudov14} for $T=0.25$K(green), $T=1$K(blue) and the temperature range $T=2.1-30.1$K(black). Red squares depict the minima ($\rho^{*},B^{*}$) for each curve. Red line demonstrate the slope $\gamma=0.57$ of typical magnetoresistivity curve(black bold) at $T=11.1$K. b) Temperature dependencies of the total $\rho(T)$ and Drude $\rho_{D}(T)$ resistivity at zero magnetic field. The T-dependent magnetoresistivity minima $\rho^{*}(T)$. Inset: Temperature dependent magnetic field of the magnetoresistivity minima $B^{*}(T)$(open squares) and the crosscheck curve(thin line) calculated using the Eq.(\ref{B_star}).}
\end{center}
\end{figure}
It is worth noting that a significant drawback of the proposed method is the assumption of precisely known values of the widths of 2D channels. In practice, the channel may contain parasitic large-scale defects that are the result of the growth of the 2D structure. In addition to the reduction in 2D mobility, large-scale defects results in uncontrolled narrowing of the channel width compared to the nominal one. In this case, the proposed method for extracting the scattering length becomes inapplicable.

\begin{figure}[tbp]
\begin{center}\leavevmode
\includegraphics[width=0.8\linewidth]{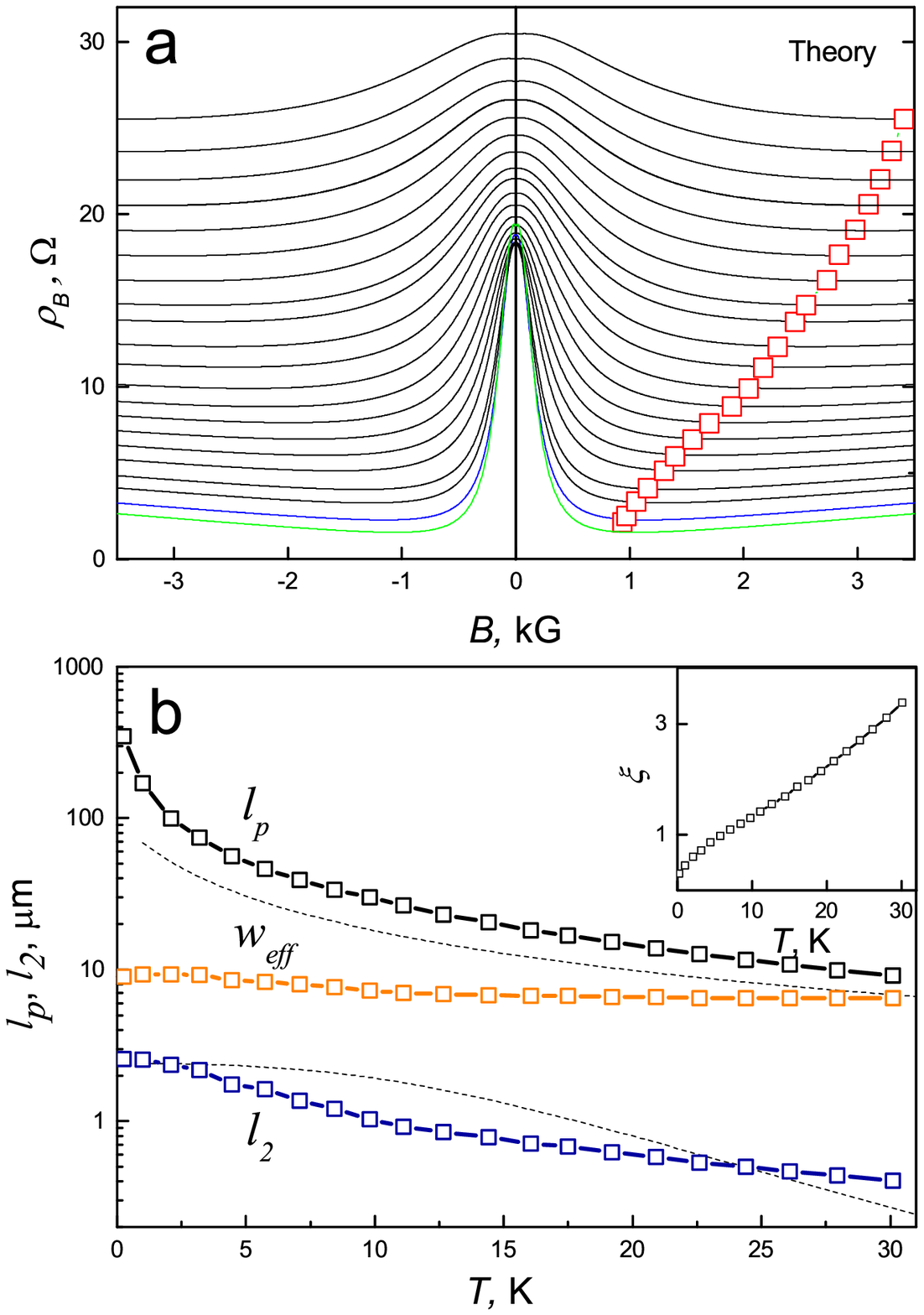} \caption[]{\label{Fig3} a) The result of calculations provided the best fit of the resistivity data in Fig.\ref{Fig2}a. b) The temperature dependent scattering lengths $l_{2}(T),l_{p}(T)$, the effective width $w_{eff}(T)$ and Gurzhi parameter $\xi(T)$(inset). Dotted lines depict the result of Ref.\cite{Alekseev16}.}
\end{center}
\end{figure}

\section{Magnetotransport in a single channel.}
\label{Magnetotransport in a single channel. Easy to make, difficult to measure}
Now consider a 2D fluid in a channel of fixed width placed in a perpendicular magnetic field. Recall that with increasing magnetic field, the viscosity of a 2D liquid decreases. In fact, the magnetic field destroys the hydrodynamic flow regime in favor of the ordinary Drude transport scenario. Usually, at a fixed temperature the magnetoresistivity first decreases and then grows upon enhancement of the magnetic field. The minimum of magnetoresistivity exists. This is precisely the behavior that can be seen in Fig.\ref{Fig2}a reproduced from Ref.\cite{Zudov14}.

Next, we present a self-consistent procedure that will allow us to reproduce the resistance curves shown in Fig.\ref{Fig2}a and, in addition, to extract the scattering lengths $l_{p},l_{2}$. First, we find from Fig.\ref{Fig2}a, and then we plot in Fig.\ref{Fig2}b the resistivity as a function of temperature at zero magnetic field, $\rho(T)$. For each curve in Fig.\ref{Fig2}a, corresponding to a certain temperature, we find the minimum of magnetoresistivity with certain values $\rho^{*},B^{*}$. The analysis of the entire set of the magnetoresistivity curves results in dependence $\rho^{*}(B^{*})$ shown by open squares in Fig.\ref{Fig2}a. Finally, the auxiliary temperature dependencies $\rho^{*}(T)$ and $B^{*}(T)$ are plotted on the main panel and in the inset to Fig.\ref{Fig2}b, respectively.

It is noteworthy that all magnetoresistivity curves in Fig.\ref{Fig2}a show a linear increase in strong magnetic fields, where hydrodynamic effects are presumably suppressed. Evidently, the positive magnetoresistivity is owned to the Drude transport regime. For the set of curves in Fig.\ref{Fig2}a we estimate the average magnetoresistivity slope as $\gamma=d\rho_{B}/dB=0.57\Omega$/kG. Therefore, it is convenient to represent Drude resistivity $\rho_{D}(T,B)$ as it follows
\begin{equation}
\rho_{D}(T,B)=\rho_{D}+\gamma B.
\label{Drude_appoximation}\\
\end{equation}
We recall that the transition from the hydrodynamic to the Drude transport regime occurs when the carriers viscosity is suppressed, when $l_{2}/r_{c}\gg 1$. The Gurzhi parameter reads $\xi_{B} \simeq 2\xi l_{2}/r_{c}=aB$. The expansion of the Eq.(\ref{Gurzhi_Resistivity}) gives the following expression
\begin{equation}
\rho_{B}=(\rho_{D}+\gamma B)(1+1/aB),
\label{Gurzhi_Resistivity_simple}\\
\end{equation}
valid in the vicinity of the magnetoresistivity minimum. A simple analysis of the Eq.(\ref{Gurzhi_Resistivity_simple})
gives the magnetic field $B^{*}=\sqrt{\frac{\rho_{D}}{a\gamma}}$ of minimum magnetoresistivity. The latter can be re-written as it follows:
\begin{equation}
B^{*}=\left ( \frac{eR_{\textrm{k}}}{2\alpha \gamma} \right )^{1/2}\cdot \frac{\sqrt{\xi}}{w},
\label{B_star}\\
\end{equation}
where $R_{\textrm{k}}=h/e^{2}$ is the Klitzing constant, then $\alpha=1/137$ is the conventional fine-structure constant.

Substituting $B^{*}$ into the Eq.(\ref{Gurzhi_Resistivity_simple}) and taking into account the equality $\rho^{*}=\rho_{B}(B^{*})$, we obtain the important relationship
\begin{equation}
\rho_{D}=\frac{\rho^{*}}{2}-\gamma B^{*}+\sqrt{\left(\frac{\rho^{*}}{2}\right )^{2}-\rho^{*}\gamma B^{*}}.
\label{Zero-field Drude}\\
\end{equation}
Thus, we have related the zero-field Drude resistivity $\rho_{D}$ with the parameters $\rho^{*},B^{*}$ of the magnetoresistivity extremum. As expected, for zero slope $\gamma=0$ we obtain $\rho_{D}\equiv\rho^{*}$ for an infinitely large magnetic field.

Using the experimental dependencies $\rho^{*}(T), B^{*}(T)$ presented in Fig.\ref{Fig2}b for magnetoresistivity minima, and the Eq.(\ref{Zero-field Drude}), we calculate and, then plot in Fig.\ref{Fig2}b the zero-field Drude resistivity $\rho_{D}(T)$. This is the central result of the present method which allows the separation of the Drude and viscous components of the total resistivity at $B=0$. The obtained dependence $\rho_{D}(T)$ exhibits nearly linear behavior caused by scattering of carriers by static defects and acoustic phonons. For the carrier density $n=2.8 \cdot 10^{11}$cm$^{-2}$, indicated for GaAs/AlGaAs 2D system\cite{Zudov14}, we compute and, then plot in Fig.\ref{Fig3},b the temperature dependence of the carriers mean free path $l_{p}(T)$. We estimate the carrier mobility $1.4 \cdot 10^{7}$cm$^{2}$B/s at $T=1$K being an order of magnitude higher than that $\sim 10^{6}$cm$^{2}$B/s reported in Ref.\cite{Zudov14} on the base of total resistivity value of $20 \Omega$ at $B=0$. For comparison with previous studies, we also placed on Fig.\ref{Fig2}b the result of Ref.\cite{Alekseev16} followed from empirical formulae on the base of Eq.(\ref{Gurzhi_Resistivity}). Note that our rigorous approach gives the correct scattering length which exceeds that reported in Ref.\cite{Alekseev16}.

Recall that zero-field resistivity $\rho(T)$ shown in Fig.\ref{Fig2},b. is described by the Eq.(\ref{Gurzhi_Resistivity}) under condition $\xi_{B}=\xi$. It is easy to solve the transcendental Eq.(\ref{Gurzhi_Resistivity}) and find the Gurzhi parameter in zero field as a function of temperature. The resulting dependence $\xi(T)$ is plotted in Fig.\ref{Fig3}b,inset.

Let us now reproduce the entire set of experimental curves shown in Fig.\ref{Fig2}a. To do this, we represent the Gurzhi parameter in a magnetic field in the form
\begin{equation}
\xi_{B}=\xi\sqrt{1+\left (\frac{2w^{2}}{r_{c} l_{p}\xi^{2}}\right )^{2}}.
\label{Gurzhi_Number}\\
\end{equation}
\begin{figure}[tbp]
\begin{center}\leavevmode
\includegraphics[width=1.0\linewidth]{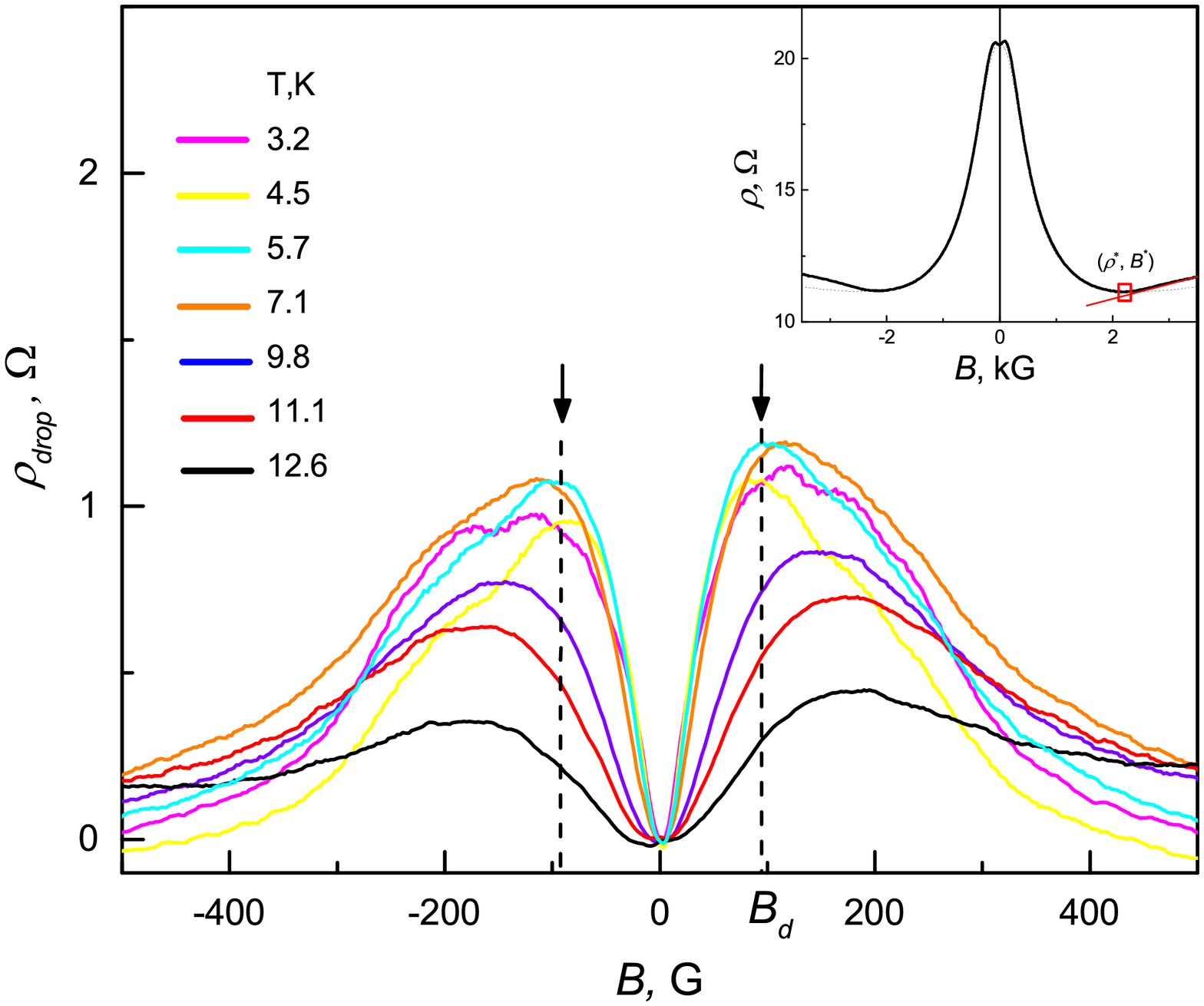} \caption[]{\label{Fig4} The droplets scattering magnetoresistance $\rho_{drop}$ calculated as the difference (see example in the inset for T=11.1K) between the initial and calculated data from Fig.\ref{Fig2}a and Fig.\ref{Fig3}a, respectively.}
\end{center}
\end{figure}
Eqs.(\ref{Gurzhi_Resistivity}),(\ref{Gurzhi_Number}),(\ref{Drude_appoximation}) in combination with the scattering lengths $l_{p}(T)$,$\xi(T)$ and the expression for the cyclotron radius $r_{c}[\mu \textrm{m}]=0.88/B$[kG] allow us to approximate each experimental curve present in Fig.\ref{Fig2}a. Our best result is depicted in Fig.\ref{Fig3}a. The inset to Fig.\ref{Fig4} demonstrates an example of the fit quality. It is important to note that each curve requires a \emph{single} value of the effective width $w_{eff}$ to fit. The resulting temperature dependence $w_{eff}(T)$ is plotted in Fig.\ref{Fig3}b. In the temperature range under study, the effective width lies within $6\div 9\mu m$ range, which is significantly less than the nominal channel size of 200$\mu m$\cite{Zudov14}. Although puzzling, this find has a simple explanation.

The narrowing of the channel width is always associated\cite{Bockhorn14} with the presence of obstacles in the form of gallium droplets, which are inevitable during the growth of a GaAs/AlGaAs heterostructure. It is assumed\cite{Alekseev16} that the average distance between drops, $d$, can determine the effective width of the hydrodynamic flux. Let us confirm this idea by analyzing two symmetrical maxima ($"\downarrow"$) visible in Fig.\ref{Fig2}a at magnetic fields $\sim \pm 100$G and resembling a "Chicago hat". We attribute these peaks to the presence of additional diffusion scattering at the edges of droplets, similar to scattering at the walls of a Hall sample known in the literature\cite{Thornton89}. At a given temperature, one can select the native and calculated magnetoresistivity curves shown in Fig.\ref{Fig2}a and Fig.\ref{Fig3}a, respectively. The difference between the two curves can be called as the "droplet" magnetoresistivity $\rho_{drop}(B)$. The temperature set of the "droplet" curves is plotted in Fig.\ref{Fig4}. It is noteworthy that the position of the "Chicago hat" peaks determines the characteristic magnetic field $B_{d}=\pm 96$ which corresponds to the carriers cyclotron radius $r_{c}=9.2\mu m$. It is obvious that droplets carrier scattering is effective when the cyclotron radius coincides with the average distance between droplets $d$ or, in turn, the effective channel width $w_{eff}\sim 8\mu m$. A simple estimate of the droplet density gives a value of $(\pi d^{2}/4)^{-1}=1.5\cdot 10^{5}$cm$^{-2}$, comparable to the value of $\sim 7.7 \cdot 10^{4}$cm$^{-2}$ reported in Ref.\cite{Bockhorn14} for the most disordered samples.

Let us return to the temperature-dependent lengths $w_{eff}(T), l_{p}(T)$ shown in Fig.\ref{Fig3}b. With the help of Gurzhi parameter $\xi(T)$ the calculation of the remaining dependence of the scattering length $l_{2}(T)$ is straightforward. The resulting curve is displayed in Fig.\ref{Fig3}b. For comparison, we placed on the same graph the result of Ref.\cite{Alekseev16}, which follows from the truncated Eq.(\ref{Gurzhi_Resistivity}) and extra assumptions about scattering rates. Our rigorous analysis leads to a completely different functional dependence $l_{2}(T)$ compared to that reported in Ref.\cite{Alekseev16}.

\begin{figure}[tbp]
\begin{center}\leavevmode
\includegraphics[width=1.0\linewidth]{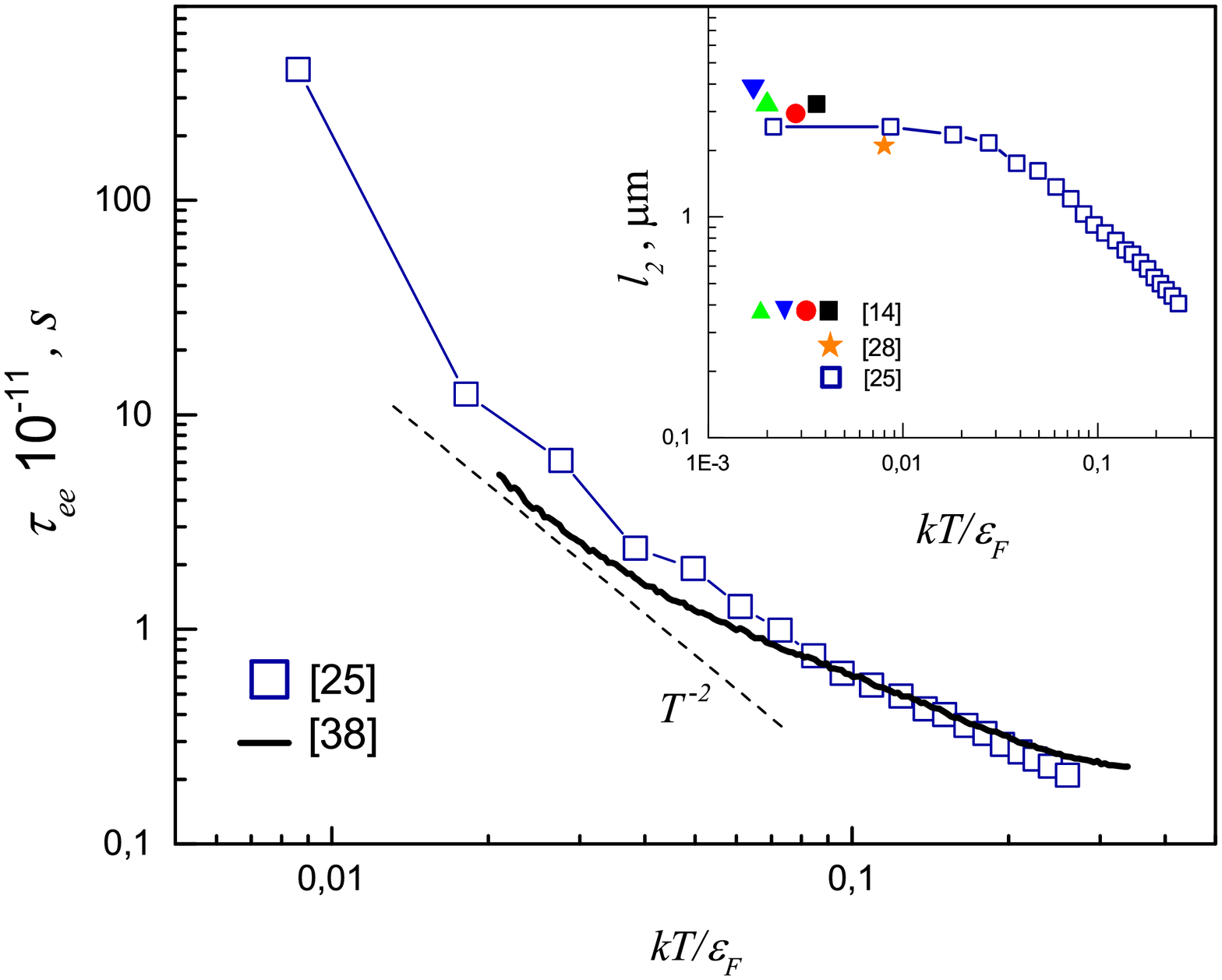} \caption[]{\label{Fig5} The e-e scattering length $\tau_{ee}(T)$ as a function of the reduced temperature compared to that reported in Ref.\cite{Sarypov24}. The dashed curve depicts the asymptote $1/T^{2}$. Inset: the scattering length $l_{2}$ vs reduced temperature reproduced from Fig.\ref{Fig3},b jointly with data\cite{Wang22,Horn21} summarized in Table.}
\end{center}
\end{figure}

Continuing the analysis, we note that the ratio of temperature to Fermi energy, $kT/\varepsilon_{F}$, would play the role of the key parameter. Accordingly, we plot in Fig.\ref{Fig5},inset the scattering length $l_{2}(T)$ as a function of this ratio. In addition, we put on the same graph the data summarized in the Table for multiple-in-a-row Hall bar samples\cite{Wang22,Horn21}. Unexpectedly, at low temperatures $kT/\varepsilon_{F}<0.01$, the scattering length $l_{2}(T)$ falls in the range of 2-3$\mu m$, regardless of the twofold variation in carrier density and mobility. We attribute the observed behavior to disorder induced contribution to length $l_{2}$ described by Eq.(\ref{Length_2}). At higher temperatures, the length $l_{2}$ decreases due to increasing the intensity of e-e interactions. 

Let us now turn to finding the e-e scattering time $\tau_{ee}(T)$. The zero-temperature extrapolation of the curve $l_{2}(T)$ in Fig.\ref{Fig5},inset gives a cutoff value $l_{2}(0)\simeq 2.56 \mu m$ owned to disorder. With the help of Eq.(\ref{Length_2}) the calculation of e-e scattering time $\tau_{ee}(T)$ is straightforward. The result is shown in Fig.\ref{Fig5}. On the same graph we have placed the data from Ref.\cite{Sarypov24} obtained by measuring the resistance of superballistic point contact. The coincidence of the both curves points to universal behavior of e-e scattering time.
Unexpectedly, at low temperatures the dependence $\tau_{ee}(T)$ is stronger than that $1/T^{2}$ suggested\cite{Pomeranchuk50} for highly degenerate electrons.

To complete the discussion of the current method, we demonstrate its self-consistency using the temperature dependence of the magnetic field.
$B^{*}$ for magnetoresistivity minima. The experimental data are displayed in Fig.\ref{Fig2}b, inset. Substituting the previously found dependencies $\xi(T), w_{eff}(T)$ into the Eq.(\ref{B_star}), we find the cross-check dependence $B^{*}(T)$ shown as a solid line in Fig.\ref{Fig2}b, inset. The coincidence of experimental and labor-intensive calculation data confirms the self-consistency of our approach.

\section{Conclusions}
\label{Conclusions}
In conclusion, we demonstrated two different methods for accurately determining carrier scattering lengths using a Hall sample with conducting channels of different widths and a single-channel samples in a perpendicular magnetic field. The analysis is based on the use of the Gurzhi model with no-slip boundary conditions. The correctness of the applicability of the boundary conditions of no-slip has been proven. Both suggested methods allow accurate determination of the the carrier mobility and viscosity. The e-e scattering time is extracted and compared to that provided by transport measurements of superballistic point contact. At low temperatures the e-e scattering time demonstrates a stronger dependence compare to $1/T^{2}$ behavior predicted by theory. We propose both developed methods as a powerful tool for viscometry and mobility assessment of 2D fluids.

\section{Acknowledgments}
\label{Acknowledgments}
The author thanks P.~S. Alekseev and A.~P. Dmitriev for useful discussions.

\bibliography{GURZHI_FLOW}

\end{document}